\documentstyle[epsf,twoside,fleqn,espcrc2]{article}

\newcommand{\AmS}{{\protect\the\textfont2
  A\kern-.1667em\lower.5ex\hbox{M}\kern-.125emS}}

\def\GeV{\mathord{\rm \;GeV}}

\def\NDR{\mathord{\rm \;NDR}}

\title{B-parameters of 4-fermion operators from lattice QCD}

\author{Rajan Gupta\address{Group T-8, MS B-285, Los Alamos National Laboratory, %
        Los Alamos, New Mexico, 87545, USA}%
        \thanks{Work supported by DoE Grand Challenges award.}
}
       
\begin{document}

\begin{abstract}
This talk summarizes the status of the calculations of $B_K$, $B_7$,
$B_8$, and $B_s$, done in collaboration with T. Bhattacharya,
G.~Kilcup, and S.~Sharpe. Results for staggered, Wilson, and
Clover fermions are presented.
\end{abstract}

\maketitle

\section{INTRODUCTION}

Reliable estimates of the matrix elements of 4-fermion operators
between hadronic states are essential in order to quantify strong
interaction corrections to weak processes.  Here we report on
calculations of mixing elements of (i) $\Delta s = 2$ operators
between $K^0$ and $\bar{K^0}$ states ($B_K$) that arise in the
calculation of the CP violation parameter $\epsilon$. The value of
$B_K$ is an essential input in pinning down the Wolfenstein parameters
$\rho$ and $\eta$ in the CKM matrix. (ii) The strong and
electromagnetic penguin operators needed to predict
$\epsilon'/\epsilon$ ($B_6$ and $B_8$).  In particular, we consider
$B_8$, which is phenomenologically important since a smaller value
means a larger $\epsilon'/\epsilon$. (iii) The $S+P$ operators needed
in the study of the lifetime differences of B mesons ($B_s$).

\section{STAGGERED RESULTS}

Staggered fermions are the method of choice for calculating kaon
matrix elements as they respect the continuum chiral Ward identities.
Our results for $B_K, B_7, $ and $B_8$ have recently been given in
\cite{BK97stag}.  They are based on the same numerical data first
presented at LATTICE 93 by Sharpe \cite{BK93srs}.  As explained in
\cite{BK97stag}, even though the statistical quality of the data is
meager by present standards, the largest source of error in the
quenched staggered theory is a systematic one -- the dependence of
$B_K$ on the lattice operator. The second new feature of the analysis,
compared to \cite{BK93srs}, is a better understanding of the matching
between lattice and continuum operators using tadpole improved 1-loop
perturbation theory, $i.e.$ the horizontal matching explained in
\cite{BK96WIL}.

\begin{figure}[t]
\vspace{9pt}
\hbox{\hskip15bp\epsfxsize=0.9\hsize \epsfbox {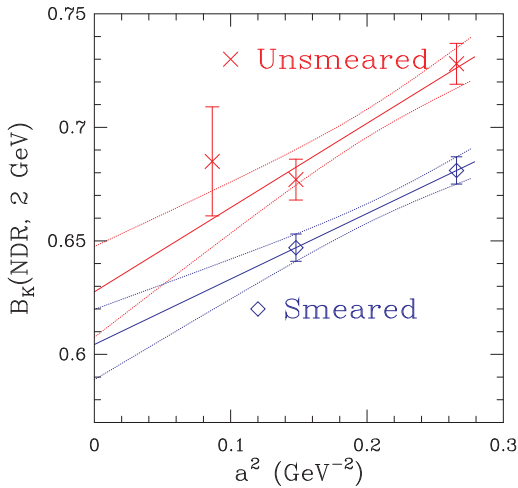}}
\vskip -0.6cm
\caption{The data for $B_K(\NDR,2\GeV)$ as a function of lattice spacing $a^2$, 
along with a linear extrapolation to $a=0$, for the smeared and unsmeared operators.}
\label{f:Bkstag}
\end{figure}

The data for both ``smeared'' ($\beta=6.0, 6.2$) and ``unsmeared''
($\beta = 6.0, 6.2, 6.4$) operators are shown in
Fig.~\ref{f:Bkstag}. (See \cite{BK97stag} for definition of these
operators.)  The extrapolation to $a=0$ is done assuming that only the
leading correction $O(a^2)$ contributes (the absence of the linear
$O(a)$ term is expected theoretically \cite{BK93srs}, and has been confirmed
numerically \cite{BK97JLQCD}).  The results
are (for $q^* = 1/a$)
\begin{eqnarray}
B_K(NDR, 2 \GeV) &=& 0.63(2) \quad {\rm unsmeared} \nonumber \\
B_K(NDR, 2 \GeV) &=& 0.60(2) \quad {\rm smeared} 
\end{eqnarray}
For the final value we take the mean
\begin{equation}
B_K(NDR, 2 \GeV) = 0.62 \pm 0.02 \pm 0.02
\label{eq:BK}
\end{equation}
where the second error covers the spread due to the operator dependence. 

The difference between the central values for smeared and unsmeared
operators is $\sim 0.025$. While our data is certainly not good enough
to argue that this difference is significant, a difference of similar
size has been reported by JLQCD \cite{BK97JLQCD}.  This difference is
an artifact of keeping only an $a^2$ correction term in the $a=0$
extrapolation as shown by the following argument. Consider two
discretizations, ${\cal O}_1$ and ${\cal O}_2$, of any 4-fermion
operator.  Let the typical lattice momenta $q^*$ associated with their
lattice measurement be $K_1/a$ and $K_2/a$ respectively.  Then, using 
the ``horizontal'' matching to the continuum scheme defined in
\cite{BK96WIL}, the results at $\mu = K_1/a$ are related as
\begin{eqnarray}
{\cal O}_1 (\mu) &=& {\cal O}_2 (\mu) 
                     ( \alpha(K_2/a)/\alpha(K_1/a))^{-\gamma_0/2\beta_0} \nonumber \\
                 &+& \left\{ 1 + X \alpha^2(q*) + Y a^2 + \ldots \right\} \ .
\end{eqnarray}
Since the factor $\alpha(K_2/a)/\alpha(K_1/a) \to 1$ as $a \to 1$, the 
two operators should give the same result in the continuum limit provided 
the extrapolation is done including factors of both $O(a)$ and $O(\alpha)$. 

To convert the quenched result in Eq.~\ref{eq:BK} to the
renormalization group invariant quantity $\widehat B$, one can proceed in
two ways: use the $n_f=0$ or the $n_f=4$ values for $\alpha_s,
\beta_0, \beta_1, \gamma_1$. The quenched result, using $\alpha(2\GeV)=0.19$, is
\begin{equation}
\widehat{B}_K  = 0.86 \pm 0.03 \pm 0.03
\end{equation}
Interestingly, using $n_f=4$ values for $\beta_0, \beta_1, \gamma_1$
and $\alpha_{\overline{MS}}^{(4)}(2\GeV)=0.3$ also gives the same
value. However, there is an uncertainty of $\sim 0.05$ in such a
conversion.  What we really predict is the quenched result
Eq.~\ref{eq:BK}.

\subsection{Electromagnetic penguins}

The quantity that enters in the standard model calculation of
$\epsilon'/\epsilon$ is $B_8$ evaluated at say $\mu=2$ GeV
\cite{BURAS97}. To get this from lattice simulations one needs to
calculate the matrix elements of both $O_7$ and $O_8$ as these mix
under a scale evolution.

Our current calculations use 1-loop matching factors for the 4-fermion
operators.  As explained in \cite{BK97stag}, if the discretization of
operators is such that $Z_P$ for the pseudoscalar bilinear is large,
then the mixing contribution due to the $P \otimes P$ term can be even
larger than the tree level result. In such a case the 1-loop
determination of the matching $Z$'s is inadequate, and no results for
$B_7$ or $B_8$ can be extracted. This is true for the unsmeared Landau
gauge operators that we have used and probably also for the gauge
invariant operators used by JLQCD \cite{BK97JLQCD} as they have the 
same $Z_P$.  On the other
hand, the 1-loop perturbative value for $Z_P$ for the smeared
operators is much smaller, and consequently results are independent of
$q^*$ to within $10\%$ as shown in Table \ref{tab:B7B8}.  (The failure
of 1-loop $Z$'s for the unsmeared operators shows up as a large
dependence of $B$-parameters on $q^*$; for example the results even
change sign between $q^* = \pi/a$ and $1/a$ \cite{BK97stag}.)

\begin{table*}[thb]
\setlength{\tabcolsep}{1.5pc}
\newlength{\digitwidth} \settowidth{\digitwidth}{\rm 0}
\catcode`?=\active \def?{\kern\digitwidth}
\begin{tabular*}{\textwidth}{@{}l@{\extracolsep{\fill}}rrrr}
\hline
Operator        & $q^*$         & $\beta=6.0$   & $\beta=6.2$   & $a=0$\\
\hline
$B_7^{3/2}$      & $1/a$         & 0.989(05)     & 0.823(16)     & 0.62(3)\\
$B_7^{3/2}$      & $\pi/a$       & 1.085(06)     & 0.903(14)     & 0.67(3)\\
$B_8^{3/2}$      & $1/a$         & 1.240(06)     & 1.030(16)     & 0.77(4)\\
$B_8^{3/2}$      & $\pi/a$       & 1.288(06)     & 1.076(17)     & 0.81(4)\\
\hline
\end{tabular*}
\caption{Results for $B_7^{3/2}(\NDR,2\GeV)$ and $B_8^{3/2}(\NDR,2\GeV)$,
at the physical kaon mass, using smeared operators. The last column gives 
the result of linear extrapolation in $a^2$.}
\label{tab:B7B8}
\end{table*}

The shaky part of this analysis is that the data at two values of $a$
are extrapolated using just the lowest order ($a^2$) correction.
Since these correction are large, further checks of these first results are
needed.

\section{WILSON FERMIONS}

Our results with Wilson fermions are exploratory.  They have been
obtained at just $\beta=6.0$, albeit on large lattices and
with high statistics \cite{BK96WIL}. The goal has been to understand
systematic errors, in particular the question of bad chiral behavior
of matrix elements induced by the mixing with wrong chirality
operators.

The general form of the kaon matrix elements, as predicted by $\chi$PT, is 
\begin{eqnarray*}
{\left\langle \overline{K^0}(p_f)
\right| {\cal O} \left| K^0(p_i) \right\rangle
\over   (8/3) f_{K,{\rm phys}}^2} =
\alpha + \beta m_K^2 + \gamma \, p_i\cdot p_f  +  \\
\qquad \delta_1 m_K^4 + \delta_2 m_K^2 p_i\cdot p_f + \delta_3 (p_i\cdot p_f )^2 + \ldots .
\label{eq:Bkcpt}
\end{eqnarray*}
where we ignore chiral logarithms and terms proportional to $(m_s-m_d)^2$.
The former are difficult to distinguish numerically from the
terms we include, while the latter we expect to be small, especially
for the range of quark masses studied.

For $B_K$, chiral symmetry predicts that $\alpha$, $\beta$, and
$\delta_1$ are zero. With Wilson fermions, the mixing with wrong
chirality operators generates these terms and, in addition, the
allowed terms $\gamma, \delta_2, \delta_3$ get contributions that have
to be eliminated. With 1-loop improved operators these artifacts are
$O(\alpha_s^2)$, but nevertheless overwhelm the signal.  Our approach
is to first remove $\alpha, \beta, \delta_1$ by studying the momentum
dependence, and secondly, roughly estimate the artifacts in $\gamma,
\delta_2, \delta_3$ using the fact that $\alpha, \beta, \delta_1$ have
to be zero. Putting all these together gives, without extrapolation in
$a$, \cite{BK96WIL}
\begin{equation}
B_K(NDR, 2 \GeV) = 0.74 \pm 0.04 \pm 0.05
\end{equation}
where the second error is an estimate of the residual contamination due
to bad chiral behavior. 

Results for the electromagnetic penguins operators, in the 
NDR scheme at $\mu=2\GeV$,  are \cite{BK96WIL}
\begin{eqnarray}
  B_7^{3/2} &=&  0.58 \pm 0.02 ({\rm stat})
                                {+0.07 \atop -0.03} ({\rm pert}) \,,  \\
  B_8^{3/2} &=&  0.81 \pm 0.03 ({\rm stat})
                                {+0.03 \atop -0.02} ({\rm pert}) \,.
\end{eqnarray}
The ``perturbative error'' reflects the dependence of the results on
the choice of $\alpha_s$ used in the matching of continuum and lattice
operators, and is comparable to or larger than the statistical errors.
A recent calculation of these by the APE collaboration using
non-perturbative matching coefficients and the $C_{SW}=1$ clover
action suggests that the errors in the 1-loop mixing coefficients may
be far more severe \cite{LAT97vladikas}.

The final quantities we consider are $B_S \equiv B_4^+$
and the related parameter $B_5^+$ as defined in \cite{BK96WIL}. 
The matrix elements we require are for $\bar{b} s$ mesons. The best 
we can do with present data is to give the result for $m_b \sim m_c$, 
\begin{eqnarray}
  B_4^{+}(\NDR,1/a) &=&  0.80 \pm 0.01 ({\rm stat}) \,,  \\
  B_5^{+}(\NDR,1/a) &=&  0.94 \pm 0.01 ({\rm stat}) \,.
\end{eqnarray}
A second limitation is that these results are at
$\mu=1/a=2.33 \GeV$, the scale at $\beta=6.0$, because the two-loop
anomalous dimension matrix needed to run to $2\;$GeV has not been
calculated.

\section{CLOVER FERMIONS}

The analysis of data with clover fermions is preliminary. The same 170
lattices used in the study with Wilson fermions \cite{BK96WIL}
are analyzed with tree-level tadpole improved clover action ($C_{SW}
= 1.4785$).  At this point we have a few qualitative statements about the 
data.

The statistical fluctuations in the matrix elements of $\cal S, \cal
P, \cal A, \cal V, \cal T$ operators are much larger compared to those
with Wilson fermions.  Curiously, these fluctuations cancel in the
five operators $O_1^+ \ldots O_5^+$ discussed above.

The 1-loop mixing factors for $O_7^{3/2}$ and $O_8^{3/2}$ are too
large and the calculation fails. The reason is the increase in $Z_P$
with $C_{SW}$.

The dominant artifacts in $B_K$, $\alpha$ and $\beta$, are roughly a
factor of five smaller compared to Wilson fermions.  As a result, the
lattice value of $B_K$ using the 1-loop improved operator improves
from $-0.30$ with Wilson fermions to $0.50$ with clover. This suggests
that a large part of chiral violations is an $O(a)$ effect.

\end{document}